\documentclass[aps,twocolumn,showpacs,superscriptaddress,prl]{revtex4}
\usepackage{graphicx,amsmath,color}
\usepackage{mathtools}

\begin{document}

\title{Chiral Supersolid in Spin-Orbit-Coupled Bose Gases with Soft-Core Long-Range Interactions}
\author{Wei Han}
\affiliation{Key Laboratory of Time and Frequency Primary Standards, National Time Service Center, Chinese Academy of Sciences, Xi'an 710600, China}
\affiliation{School of Astronomy and Space Science, University of Chinese Academy of Sciences, Beijing 100049, China}
\author{Xiao-Fei Zhang}
\email{xfzhang@ntsc.ac.cn}\affiliation{Key Laboratory of Time and Frequency Primary Standards, National Time Service Center, Chinese Academy of Sciences, Xi'an 710600, China}
\affiliation{School of Astronomy and Space Science, University of Chinese Academy of Sciences, Beijing 100049, China}
\author{Deng-Shan Wang}
\affiliation{School of Science, Beijing Information Science and Technology University, Beijing 100192, China}
\author{Hai-Feng Jiang}
\affiliation{Key Laboratory of Time and Frequency Primary Standards, National Time Service Center, Chinese Academy of Sciences, Xi'an 710600, China}
\affiliation{School of Astronomy and Space Science, University of Chinese Academy of Sciences, Beijing 100049, China}
\author{Wei Zhang}
\email{wzhangl@ruc.edu.cn}\affiliation{Department of Physics, Renmin University of China, Beijing 100872, China}
\author{Shou-Gang Zhang}
\email{szhang@ntsc.ac.cn}\affiliation{Key Laboratory of Time and Frequency Primary Standards, National Time Service Center, Chinese Academy of Sciences, Xi'an 710600, China}
\affiliation{School of Astronomy and Space Science, University of Chinese Academy of Sciences, Beijing 100049, China}

\begin{abstract}
Chirality represents a kind of symmetry breaking characterized by the noncoincidence of an object with its mirror image and has been attracting intense attention in a broad range of scientific areas. The recent realization of spin-orbit coupling in ultracold atomic gases provides a new perspective to study quantum states with chirality. In this Letter, we demonstrate that the combined effects of spin-orbit coupling and interatomic soft-core long-range interaction can induce an exotic supersolid phase in which the chiral symmetry is broken with spontaneous emergence of circulating particle current. This implies that a finite angular momentum can be generated with neither rotation nor effective magnetic field. The direction of the angular momentum can be altered by adjusting the strength of spin-orbit coupling or interatomic interaction. The predicted chiral supersolid phase can be experimentally observed in Rydberg-dressed Bose-Einstein condensates with spin-orbit coupling.
\end{abstract}

 \maketitle

\textit{Introduction.---}Chirality is a universal and fascinating phenomenon in nature~\cite{McGuire,Yoon}. Exploring and investigating new states of matter with chirality is a prominent subject in physics, and can light the way to a deeper understanding of nature and provide clues for designing novel functional
materials. The recent discovery of exotic chiral matter involving chiral superconductors~\cite{Kallin}, chiral electrons~\cite{Dai}, chiral domain walls~\cite{Parkin,Beach,Schmid}, and chiral skyrmions~\cite{Tokura} has attracted extensive interests of physicists. In many of these systems, the existence of spin-orbit (SO) coupling plays an important role in the symmetry breaking of chirality. Recent experimental realization of SO coupling in ultracold quantum gases~\cite{Spielman,Jing-Zhang,Zwierlein,Jing-Zhang2,Shuai-Chen} provides a highly controllable platform for the study of chirality~\cite{JHHan,SSZhang}.

Even though there are plenty of studies of SO coupling, most existing works focus only on hard-core systems, where the interatomic interaction is manifested as zero-range contact~\cite{Galitski,Dalibard,Goldman,Hui-Zhai,Hui-Zhai2,Yirev,Jing-Zhang3,Congjun-Wu,Congjun-Wu2} or long-range dipolar potentials~\cite{Su-Yi,Clark,Demler}. However, soft-core interaction can also be realized in Bose gases with Rydberg dressing technology~\cite{Pohl,WCWu,Heidemann}. The essential difference between the hard-core and soft-core interactions is the behavior of potential when two atoms are brought to close distance. For the hard-core case, the interaction tends to infinity, while for soft-core systems, the interaction potential tends to a finite value. Previous investigations~\cite{Boninsegni,Boninsegni2} suggested that the soft-core long-range interaction can induce a spontaneous supersolid, which is a long-sought exotic phase that behaves simultaneously as a solid and a friction-free superfluid~\cite{Balibar,Andreev,Chester,Leggett,Kim}. With the new ingredient of SO coupling, an intriguing question is can we have a supersolid phase with chiral symmetry breaking in a Bose gas with soft-core long-range interactions.

In this Letter, we investigate the ground-state quantum phases of Bose gases with SO coupling and soft-core long-range interactions. A surprising finding is that the combined effects of SO coupling and soft-core long-range interaction can lead to a chiral supersolid in which spontaneous circulation of particles emerges in each unit cell. This implies that a finite angular momentum is generated by the chirality imposed by SO coupling, which is in stark contrast to the general expectation that the ground state of a many-body system cannot possess finite total angular momentum~\cite{Bohm,Momoi} and also goes beyond the traditional means of yielding angular momentum by external rotation~\cite{Dalibard2,Ketterle} or synthetic magnetic fields~\cite{Spielman2}. The direction of the angular momentum is associated with the form of phase separation and can be altered by adjusting the strength of SO coupling or interatomic interaction. In addition, it is revealed that the Rashba and Dresselhaus SO couplings lead to opposite chiralities of the particle currents.

\textit{Model.---}We consider a homogeneous two-dimensional SO-coupled Bose-Einstein condensate (BEC) with soft-core long-range interactions. The Hamiltonian reads in the Gross-Pitaevskii mean-field approximation as
\begin{eqnarray}
&&\mathcal{H}=\int\!\! d\mathbf{r}\mathbf{\Psi }^{\dag }\!\left(\!-\frac{\hbar ^{2}%
\boldsymbol{\nabla }^{2}}{2M}+\mathcal{V}_{\text{SO}}\!\right)\!\mathbf{\Psi }\notag \\
&&+\frac{1}{2}\!\int\!\! d\mathbf{r}\sum\limits_{i,j=\uparrow ,\downarrow
}g_{ij}\!\Psi _{i}^{\ast }\!\left( \mathbf{r}\right)\!\Psi
_{j}^{\ast }\!\left( \mathbf{r}\right)\!\Psi _{j}\!\left( \mathbf{r}%
\right)\!\Psi _{i}\!\left( \mathbf{r}\right)\notag \\
&&+\frac{1}{2}\!\int\!\! d\mathbf{r} d\mathbf{r^{\prime }}\!\!\!\!\sum\limits_{%
i,j=\uparrow ,\downarrow}\!\Psi _{i}^{\ast }\!\left( \mathbf{r}%
\right)\!\Psi _{j}^{\ast }\!\left( \mathbf{r^{\prime }}\right)\!U_{ij}\!\left( \mathbf{r\!-\!r^{\prime }}\right)\!\Psi _{j}\!\left( \mathbf{%
r^{\prime }}\right)\!\Psi _{i}\!\left( \mathbf{r}\right)\!,\quad\label{Model Hamiltonian}
\end{eqnarray}
where the spinor order parameter $\mathbf{\Psi}=[\Psi_{\uparrow}(\mathbf{r}),\Psi_{\downarrow}(\mathbf{r})]^\top$ with $\mathbf{r}=(x,y)$ and is normalized to satisfy $\int d\mathbf{r} \mathbf{\Psi}^{\dag}\mathbf{\Psi}=~N$. The SO coupling term can be written as
$\mathcal{V}_{\text{SO}}=-i\hbar\kappa(\sigma_{x}\partial_{x}\pm\sigma_{y}\partial_{y})$,
where $\sigma_{x,y}$ are the Pauli matrices, and $\kappa$ denotes the SO coupling strength. Here, the sign $``\pm"$ distinguishes the types of SO couplings as Rashba for $``+"$ and Dresselhaus for $``-"$. The strength of the contact interaction is characterized by $g_{ij}$, and here we focus on the SU(2) symmetric case with $g=g_{\uparrow\uparrow}=g_{\downarrow\downarrow}=g_{\uparrow\downarrow}$. The effective potential describing the soft-core long-range interaction is written as $U_{ij}\left(\mathbf{r}\right)=\tilde{C}_{6}^{(ij)}/\left(R_{c}^{6}+|\mathbf{r}|^6\right)$, where $\tilde{C}_{6}^{(ij)}$ characterizes the interaction strength and $R_{c}$ represents the blockade radius~\cite{WCWu}.

The experimental realization of the model Hamiltonian in Eq.~(\ref{Model Hamiltonian}) may be achieved with the (5$S_{1/2}$, $F=1$) ground electronic manifold of $^{87}$Rb atoms, where two of the hyperfine states $|F=1,m_{F}=-1>$ and $|F=1,m_{F}=0>$ are chosen to simulate the spin-up $|\uparrow>$ and spin-down $|\downarrow>$ components, respectively~\cite{Supp}. While the contact interaction exists naturally, the soft-core long-range interaction can be artificially created by using the Rydberg dressing technique, where the ground-state atoms are weakly coupled to a highly excited Rydberg state by an off-resonant two-photon process~\cite{Pohl,WCWu,Heidemann}. The soft-core long-range interaction strengths $\tilde{C}_{6}^{(ij)}$ and blockade radius $R_{c}$ depend on the two-photon Rabi frequency and detuning, and can be tuned within a wide range in an experimentally accessible region~\cite{Supp}. The Rashba and Dresselhaus SO coupling may be created by modulating the gradient magnetic field~\cite{ZFXu,BMAnderson,Ruquan-Wang} or Raman laser dressing~\cite{Campbell}. The two-dimensional geometry can be realized by imposing a strong harmonic potential $V(z)=M\omega_{z}^2z^2/2$ along the axial direction with the characteristic length $a_{h_z}=\sqrt{\hbar/M\omega_{z}}\ll R_{c}$, in which case, the effective contact interaction strength is given by $g_{ij}=\sqrt{8\pi}(\hbar^2/M)(a_{ij}/a_{h_{z}})$ with $a_{ij}$ being the $s$-wave scattering length~\cite{Yefsah}.

\begin{figure}[t]
\centerline{\includegraphics[width=8cm,clip=]{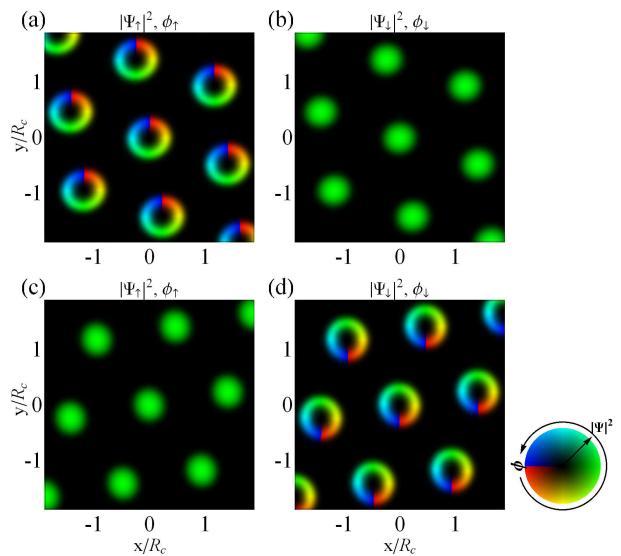}}
\caption{Chiral supersolid induced by Rashba spin-orbit coupling and soft-core long-range interactions. The density and phase distributions represented, respectively, by brightness and color are shown in (a) and (b) with the soft-core long-range interactions $\tilde{C}_{6}^{(\uparrow\uparrow)}N=2\tilde{C}_{6}^{(\downarrow\downarrow)}N=2500\ \hbar^2R^4_{c}/M$, and in (c) and (d) with $\tilde{C}_{6}^{(\downarrow\downarrow)}N=2\tilde{C}_{6}^{(\uparrow\uparrow)}N=2500\ \hbar^2R^4_{c}/M$. The directions of the arrows in the color wheel indicate the elevation of the respective quantities. Other parameters are fixed at $\tilde{C}_{6}^{(\uparrow\downarrow)}N=1250\ \hbar^2R^4_{c}/M$, $\kappa=4\ \hbar/MR_{c}$, and $gN=1000\ \hbar^2/M$. Here, the soft-core long-range interaction strengths $\tilde{C}_{6}^{(ij)}$ are considered in an experimentally achievable parameter range~\cite{Supp}.} \label{fig1}
\end{figure}

\textit{Chiral supersolid.---}The many-body ground states can be obtained by numerically minimizing the Hamiltonian functional given by Eq.~(\ref{Model Hamiltonian})~\cite{Supp}. In the case without SO coupling, it has been known that the soft-core long-range interactions can induce a supersolid phase with roton-type mode softening~\cite{Pohl,WCWu}. In that case, the Hamiltonian is symmetric with respect to a chiral operation ${\hat O} = {\hat K}$, where ${\hat K}$ denotes the complex conjugate. The existence of Rashba or Dresselhaus SO coupling explicitly breaks this chiral symmetry and leads to an exotic chiral supersolid phase with each unit cell possessing a clockwise or counterclockwise circulation of phase as shown in Fig.~\ref{fig1}. In this phase, the two spin components are separated along the radial direction in each unit cell. The component with weaker intracomponent interaction always lies in the center surrounded by the component with opposite spin with stronger intracomponent interaction. While the phase of the core component is trivial, there exists a $2\pi$ phase gradient along a closed path around the toroidal component forming a vortex in each unit cell. A surprising observation is that all the vortices choose the same direction of circulation, which is distinct from those observed in an SO-coupled hard-core system, where vortices and antivortices emerge in pairs~\cite{Santos,Hu,ZFXu2,SCGou,Ueda,Mottonen,Ueda2,SCGou2}.

To get a physical picture of the emergence of these aligned vortices, we rewrite the Rashba SO coupling term $\mathcal{H_{\text{so}}} = -i\hbar\kappa\int d\mathbf{r}\mathbf{\Psi }^{\dag } (\sigma_{x}\partial_{x}+\sigma_{y}\partial_{y})\mathbf{\Psi }$ in the polar coordinate~$(r, \varphi)$,
\begin{eqnarray}
\mathcal{H_{\text{SO}}}\!=\!-2\kappa\!\!\int_{\Lambda_{0}}\!\!\!\! d\mathbf{r} \text{Re}\left[\Psi _{\uparrow }^{\ast }\exp \left(
-i\varphi \right) \left( i\frac{\partial }{\partial r}+\frac{\partial }{r\partial \varphi }\right) \Psi _{\downarrow }\right]
\label{SOC Hamiltonian}
\end{eqnarray}
with $\Lambda_{0}$ defining the range of integration within one unit cell of the supersolid crystalline structure and decompose the wave function by its density and phase as $\Psi_{j}=\sqrt{n_{j}}\exp(i\theta_{j})$. For the core component (denoted by the subscript ``$\bullet$"), the phase must satisfy $\partial \theta_{\bullet}/\partial \varphi=0 $ to avoid energy dissipation. In addition, by taking into account the rotational symmetry and neglecting the radial diffusion, it is natural to assume $\partial n_{j}/\partial\varphi=0$ and $\partial\theta_{j}/\partial r=0$. As a result, we have
\begin{eqnarray}
\mathcal{H_{\text{SO}}}\!=\!2\kappa\!\!\int_{\Lambda_{0}}\!\!\!\!d\mathbf{r} \big[\sin(\theta_{\bullet}-\theta_{\circ}\pm\varphi)\sqrt{n_{\circ}}\partial_{r}\sqrt{n_{\bullet}}\big],\label{SOC Hamiltonian 2}
\end{eqnarray}
where the subscript ``$\circ$" denotes the surrounding toroidal component, and the sign ``$\pm$" takes ``$+$" (``$-$") if the core component is spin-up (spin-down). In order to minimize the SO coupling energy, it is preferred that
\begin{eqnarray}
\theta_{\bullet}-\theta_{\circ}\pm\varphi=\frac{\pi}{2}+2\pi l, (l\in Z)\label{relative phase}
\end{eqnarray}
with $\theta_{\bullet}$ a constant. This result indicates that the surrounding component tends to have a $+2\pi$ ($-2\pi$) phase gradient if the core component is spin-up (spin-down), consistent with the numerical results shown in Fig.~\ref{fig1}.

The radial phase separation within a unit cell and the nontrivial circulation of phase can be regarded as a topological spin texture. By defining the Bloch vector $\mathbf{s}=\mathbf{\Psi }^{\dag }\mathbf{\sigma}\mathbf{\Psi }/|\mathbf{\Psi }|^2$, which projects the state $\mathbf{\Psi}$ onto the surface of a unit Bloch sphere, we obtain from Eq.~(\ref{relative phase}) that $s_{x}=\pm\sqrt{1-s_{z}^{2}}\sin\varphi$, $s_{y}=\mp\sqrt{1-s_{z}^{2}}\cos\varphi$, and $s_{z}=(n_{\uparrow}-n_{\downarrow})/(n_{\uparrow}+n_{\downarrow})$. Obviously, by running over
a unit cell, the Bloch vectors cover the Bloch sphere for only one time. Thus, the chiral supersolid phase shown in Fig.~\ref{fig1} features similar properties as the Skyrmion crystal in magnetic materials~\cite{YTokura,CPfleiderer,CPfleiderer2,YTokura2,YTokura3}. This similarity can be understood from the perfect match of the Rashba-BEC Hamiltonian to that used in the study of chiral magnets with Dzyaloshinskii-Moriya interactions~\cite{JHHan}.

\begin{figure}[t]
\centerline{\includegraphics[width=8cm,clip=]{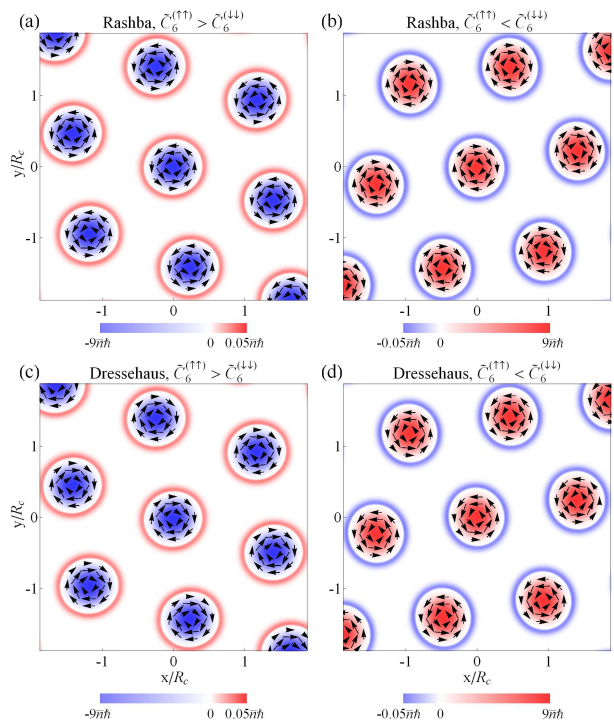}}
\caption{Particle currents $\mathbf{j}$ and longitudinal magnetizations $\mathbf{S}_{z}$ of the spin induced by (a),(b) Rashba spin-orbit coupling and (c),(d) Dresselhaus spin-orbit coupling. The color map and black arrows represent $\mathbf{S}_{z}$ and $\mathbf{j}$, respectively, where the colors ranging from blue to red describe the values from the minimum to the maximum. The parameters used are identical to those in Fig.~\ref{fig1}.} \label{fig2}
\end{figure}

The chiral supersolid acquires a spontaneous circulating particle current within each unit cell. In the hydrodynamic theory~\cite{JHHan,Baym,Fetter}, the mass conservation requires that the actual particle current in the presence of a gauge potential is given by
\begin{eqnarray}
\mathbf{j}=\frac{\hbar}{2Mi}\left[\mathbf{\Psi }^{\dag }\boldsymbol{\nabla}\mathbf{\Psi }-(\boldsymbol{\nabla}\mathbf{\Psi }^{\dag })\mathbf{\Psi }\right]-\frac{1}{M}\mathbf{\Psi }^{\dag }\mathbf{A}\mathbf{\Psi },\label{partilce current}
\end{eqnarray}
where the gauge potential is $\mathbf{A}=-\kappa M(\sigma_{x},\sigma_{y})$ for the Rashba SO coupling, and $\mathbf{A}=-\kappa M(\sigma_{x},-\sigma_{y})$ for the Dresselhaus SO coupling. While the canonical part (first term) of the particle current depends on the phase gradient $\nabla \theta_{j}$, the gauge part (second term) is related to the phase difference $\theta_{\bullet}-\theta_{\circ}$. For the Rashba case, according to the phase relation in Eq.~(\ref{relative phase}), the particle current can be expressed as
\begin{eqnarray}
\mathbf{j}_R=\frac{\hbar}{M}\frac{n_{\circ}}{r}\hat{e}_{\pm\varphi}-2\kappa\sqrt{n_{\uparrow}n_{\downarrow}}\hat{e}_{\pm\varphi},\label{circulating current}
\end{eqnarray}
where the direction $\hat{e}_{+\varphi}$ ($\hat{e}_{-\varphi}$) is counterclockwise (clockwise) if the core component is spin-up (spin-down). It can be found that the canonical and gauge parts of the particle currents always take opposite circulating directions to guarantee less energy cost. In particular, for stronger SO coupling, the gauge part will play a dominant role, and the direction of the circulating current and the spin orientation in the vortex core satisfy the left-hand rule~\cite{Note01}, as shown in Figs.~\ref{fig2}(a) and \ref{fig2}(b) obtained by numerical simulations. For the Dresselhaus-type SO coupling, a similar analysis as above leads to a particle current
\begin{eqnarray}
\mathbf{j}_D=-\frac{\hbar}{M}\frac{n_{\circ}}{r}\hat{e}_{\pm\varphi}+2\kappa\sqrt{n_{\uparrow}n_{\downarrow}}\hat{e}_{\pm\varphi}.\label{circulating current D}
\end{eqnarray}
Thus, the Dresselhaus SO coupling induces an opposite chirality of the particle current with that of the Rashba case, which is verified by numerical simulations as shown in Figs.~\ref{fig2}(c) and \ref{fig2}(d).

The generation of chiral circulating current implies that their exists a finite angular momentum in the ground state of the supersolid phase. According to Eq. (\ref{circulating current}), the angular momentum induced by Rashba SO coupling in each unit cell can be expressed as
\begin{eqnarray}
l_{z}=\pm\int_{\Lambda_{0}} d\mathbf{r}\left[\hbar n_{\circ}-2\kappa M\sqrt{n_{\uparrow}n_{\downarrow}}r\right].\label{kinetic angular momentum}
\end{eqnarray}
As all vortices circulate in the same direction, the total angular momentum of the system is nonzero. We emphasize that such an emergence of finite angular momentum is a direct consequence of broken chiral symmetry, which is in stark contrast to the traditional means of yielding angular momentum by external rotation~\cite{Dalibard2,Ketterle} or synthetic magnetic fields~\cite{Spielman2}. In addition, the direction of the angular momentum is determined by the spin orientation in the vortex core; thus, it can be altered by changing the relative strength of the intracomponent interactions $\tilde{C}_{6}^{(\uparrow\uparrow)}$ and $\tilde{C}_{6}^{(\downarrow\downarrow)}$.

The total spin angular momentum $S_{z}$ is also nonzero in the chiral supersolid. From numerical simulations, we find that most of the particles prefer to reside in the vortex core with weaker intracomponent interaction, leaving fewer particles in the surrounding ring. For the parameters we have examined, the surrounding toroidal component constitutes no more than $10\%$ particles in number~\cite{Note02}. In particular, the population of the toroidal component constitutes about $9.3\%$ in Figs.~\ref{fig1} and \ref{fig2}; thus, the total spin angular momentum is about $S_{z}=\frac{\hbar\langle \sigma_{z}\rangle}{2}\approx \pm 0.4N\hbar$. From Fig.~\ref{fig2}, we also find that the directions of the total spin angular momentum and orbital angular momentum are opposite for the Rashba SO coupling but the same for the Dresselhaus case.
\begin{figure}[t]
\centerline{\includegraphics[width=8cm,clip=]{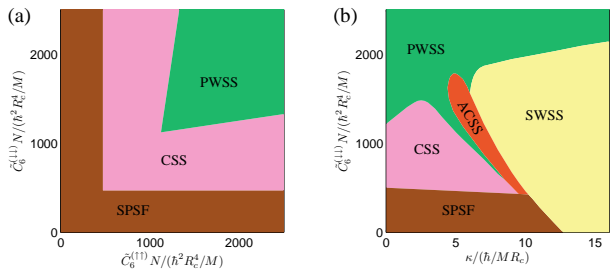}}
\caption{(a) Phase diagram by varying the soft-core long-range interaction strengths $\tilde{C}_{6}^{(\uparrow\uparrow)}$ and $\tilde{C}_{6}^{(\downarrow\downarrow)}$. (b) Phase diagram by varying the Rashba spin-orbit-coupling strength $\kappa$ and the soft-core long-range interaction strength $\tilde{C}_{6}^{(\downarrow\downarrow)}$. The spin-orbit-coupling strength is fixed at $\kappa=4\ \hbar/MR_{c}$ in (a), and the soft-core long-range interaction strength is fixed at $\tilde{C}_{6}^{(\uparrow\uparrow)}N=2500\ \hbar^2R^4_{c}/M$ in (b). Other parameters are taken as $\tilde{C}_{6}^{(\uparrow\downarrow)}N=1250\ \hbar^2R^4_{c}/M$ and $gN=1000\ \hbar^2/M$.} \label{fig3}
\end{figure}

\textit{Phase diagram.---}Next, we map out the ground-state phase diagram as a function of the soft-core long-range interaction and SO coupling strengths by running code for a grid of parameter values. In addition to the chiral supersolid (CSS) phase discussed above, two other types of supersolid phases named the plane-wave supersolid (PWSS) and the standing-wave supersolid (SWSS) are discovered as shown in Fig.~\ref{fig3}. In both the PWSS and SWSS phases, the system renders a translation symmetry breaking to form a crystalline structure, as demanded for a supersolid phase. For the PWSS phase, the local condensate wave function within each unit cell features a phase modulation along a given direction [Figs.~\ref{fig4}(a) and \ref{fig4}(b)]. For the SWSS phase, the condensate wave function is characterized by density modulation and the formation of stripes [Figs.~\ref{fig4}(c) and \ref{fig4}(d)]. Note that the local configurations of the PWSS and SWSS phases are very similar to the plane-wave and stripe phases discovered in a hard-core Bose system~\cite{Hui-Zhai3,Ketterle2,Ho,Shuai-Chen2,Stringari} and can be attributed to the competition between intra- and intercomponent interactions.

We also notice that along the diagonal line in Fig.~\ref{fig3}(a) with $\tilde{C}_{6}^{(\uparrow\uparrow)} = \tilde{C}_{6}^{(\downarrow\downarrow)}$, the Hamiltonian Eq. (\ref{Model Hamiltonian}) possesses a time reversal symmetry ${\hat T} = i \sigma_y {\hat K}$. This symmetry will be spontaneously broken in the ground state of a chiral supersolid, which randomly chooses one from the two degenerate configurations as shown in Fig.~\ref{fig1}.

In the phase diagram of Fig.~\ref{fig3}(b), we find another anomalous chiral supersolid (ACSS) phase. This phase also features chirality with finite spin and orbital angular momenta as the CSS phase discussed above. The only difference here is that the spin component with weaker intracomponent interaction prefers residing in the surrounding toroidal ring rather than the vortex core. For a conventional BEC without SO coupling, this form of phase separation is, in general, energetically unfavorable~\cite{Cornell}. In the present case with SO coupling, however, a strong SO coupling requires a large angular momentum which can be more easily accommodated by a ring with a higher number density. Notice that by tuning through the transition from the CSS to the ACSS phases with adjusting the SO coupling strength and the interatomic interaction, one can change the direction of the angular momentum as can be read from Eq.~(\ref{kinetic angular momentum}).

On the phase diagrams Figs.~\ref{fig3}(a) and \ref{fig3}(b), there also exists a spin-polarized superfluid phase if one of the intracomponent interactions $\tilde{C}_{6}^{(\uparrow\uparrow)}$ and $\tilde{C}_{6}^{(\downarrow\downarrow)}$ is very weak. In this phase, all particles condense at the component with weaker soft-core long-range interaction, and the density and phase are uniformly distributed in space.

\begin{figure}[t]
\centerline{\includegraphics[width=8cm,clip=]{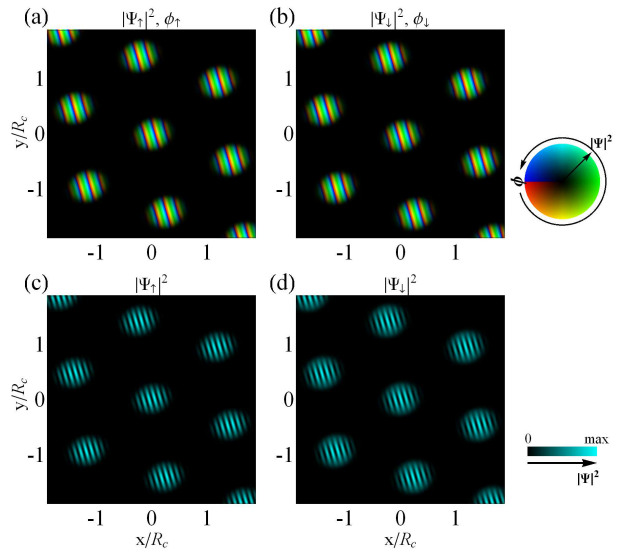}}
\caption{Configurations of the (a),(b) PWSS and (c),(d) SWSS phases. The soft-core long-range interaction strengths are $\tilde{C}_{6}^{(\downarrow\downarrow)}N=2200\ \hbar^2R^4_{c}/M$ for the PWSS phase and $\tilde{C}_{6}^{(\downarrow\downarrow)}N=1875\ \hbar^2R^4_{c}/M$ for the SWSS phase. Other parameters are taken as $\tilde{C}_{6}^{(\uparrow\uparrow)}N=2500\ \hbar^2R^4_{c}/M$, $\tilde{C}_{6}^{(\uparrow\downarrow)}N=1250\ \hbar^2R^4_{c}/M$, $\kappa=16\ \hbar/MR_{c}$ and $gN=1000\ \hbar^2/M$.} \label{fig4}
\end{figure}

\textit{Conclusion.---}In summary, we have investigated the ground-state phase diagram of spin-orbit-coupled Bose gases with soft-core long-range interactions. We have found that the system can stabilize an exotic chiral supersolid phase, which shows many unique properties: (\expandafter{\romannumeral1}) There exists a spontaneous circulating particle current in each unit cell; (\expandafter{\romannumeral2}) the system gains a finite angular momentum with neither rotation nor effective magnetic field, whose direction can be altered by adjusting the strength of spin-orbit coupling or interatomic interaction; (\expandafter{\romannumeral3}) in some parameter regions, the chiral supersolid manifests anomalous behavior of phase separation. All these aspects go beyond our existing knowledge of generating angular momentum and phase separation, and bring new perspectives on the physics of spin-orbit coupling and supersolid phenomena.

We would like to thank Hiroki Saito, Yi-Cai Zhang and Shang-Shun Zhang for helpful discussions. This work was supported by the National Key R\&D Program of China under Grant No. 2018YFA0306501, the National Natural Science Foundation of China under Grants No. 11775253, No. 11774425, No. 11772177, No. 11704383, No. 11603030, No. 11522436, No. 11547126, No. 11547194, and No. 11434011, the CAS ``Light of West China" Program under Grants No. XAB2016B73 and No. XAB2017A04, the Youth Innovation Promotion Association CAS under Grant No. 2015334, the Beijing Natural Science Foundation under Grant No. 1182009, and the Research Funds of Renmin University of China under Grant No. 16XNLQ03.

\newpage

\begin{widetext}

\section{supplementary Materials}

\noindent \textbf{Numerical details.} The many-body ground states can be obtained by numerically minimizing the Hamiltonian functional given by Eq. (1) of the main text. A widely used method for the minimization is the imaginary time algorithm~\cite{Dalfovo}, which can be reduced to solving the dimensionless evolution equations
\begin{subequations}
\begin{equation}
-\partial
_{t}\psi_{\uparrow}=\left[-\frac{1}{2}\nabla^{2}+\beta_{\uparrow\uparrow}|\psi_{\uparrow}|^{2}+\beta_{\uparrow\downarrow}|\psi_{\downarrow}|^{2}+\zeta_{\uparrow\uparrow} \int\frac{|\psi_{\uparrow}(\mathbf{r}')|^{2}}{1+|\mathbf{r}-\mathbf{r}'|^6} d\mathbf{r}'+\zeta_{\uparrow\downarrow} \int\frac{|\psi_{\downarrow}(\mathbf{r}')|^{2}}{1+|\mathbf{r}-\mathbf{r}'|^6} d\mathbf{r}'\right]\psi_{\uparrow}-\kappa\left(
i\partial _{x}+\partial
_{y}\right) \psi_{\downarrow},\tag{S1a}
\end{equation}
\begin{equation}
-\partial
_{t}\psi_{\downarrow}=\left[-\frac{1}{2}\nabla^{2}+\beta_{\downarrow\uparrow}|\psi_{\uparrow}|^{2}+\beta_{\downarrow\downarrow}|\psi_{\downarrow}|^{2}+\zeta_{\downarrow\uparrow} \int\frac{|\psi_{\uparrow}(\mathbf{r}')|^{2}}{1+|\mathbf{r}-\mathbf{r}'|^6} d\mathbf{r}'+\zeta_{\downarrow\downarrow} \int\frac{|\psi_{\downarrow}(\mathbf{r}')|^{2}}{1+|\mathbf{r}-\mathbf{r}'|^6} d\mathbf{r}'\right]\psi_{\downarrow}-\kappa\left(
i\partial _{x}-\partial _{y}\right) \psi_{\uparrow},\tag{S1b}
\end{equation}
\end{subequations}
under the constraint condition $\int\big(|\psi_{\uparrow}|^2+|\psi_{\downarrow}|^2\big) d\mathbf{r}=1$. We develop a backward-forward Euler Fourier-pseudospectral discretization, which has been previously used in a pure hard-core system with spin-orbit (SO) coupling and demonstrated to be an efficient and accurate numerical scheme~\cite{Wei-Han}. In order to get rid of the spatial derivatives, we transform Eqs.~(S1a) and (S1b) into the Fourier space
\begin{subequations}
\begin{equation}
-\partial
_{t}\tilde{\psi}_{\uparrow}=\frac{\omega_{x}^2+\omega_{y}^2}{2}\tilde{\psi}_{\uparrow}+\kappa\left(
\omega_{x}-i\omega_{y}\right) \tilde{\psi}_{\downarrow}+\tilde{G}_{\uparrow},\tag{S2a}
\end{equation}
\begin{equation}
-\partial
_{t}\psi_{\downarrow}=\frac{\omega_{x}^2+\omega_{y}^2}{2}\psi_{\downarrow}+\kappa\left(
\omega_{x}+i\omega_{y}\right) \tilde{\psi}_{\uparrow}+\tilde{G}_{\downarrow},\tag{S2b}
\end{equation}
\end{subequations}
where $\tilde{G}_{\uparrow, \downarrow}$ takes the Fourier transform of the nonlinear terms
\begin{subequations}
\begin{equation}
G_{\uparrow}=\left[\beta_{\uparrow\uparrow}|\psi_{\uparrow}|^{2}+\beta_{\uparrow\downarrow}|\psi_{\downarrow}|^{2}+\zeta_{\uparrow\uparrow} \int\frac{|\psi_{\uparrow}(\mathbf{r}')|^{2}}{1+|\mathbf{r}-\mathbf{r}'|^6} d\mathbf{r}'+\zeta_{\uparrow\downarrow} \int\frac{|\psi_{\downarrow}(\mathbf{r}')|^{2}}{1+|\mathbf{r}-\mathbf{r}'|^6} d\mathbf{r}'\right]\psi_{\uparrow},\tag{S3a}
\end{equation}
\begin{equation}
G_{\downarrow}=\left[\beta_{\downarrow\uparrow}|\psi_{\uparrow}|^{2}+\beta_{\downarrow\downarrow}|\psi_{\downarrow}|^{2}+\zeta_{\downarrow\uparrow} \int\frac{|\psi_{\uparrow}(\mathbf{r}')|^{2}}{1+|\mathbf{r}-\mathbf{r}'|^6} d\mathbf{r}'+\zeta_{\downarrow\downarrow} \int\frac{|\psi_{\downarrow}(\mathbf{r}')|^{2}}{1+|\mathbf{r}-\mathbf{r}'|^6} d\mathbf{r}'\right]\psi_{\downarrow}.\tag{S3b}
\end{equation}
\end{subequations}
The convolution in Eqs.~(S3a) and (S3b) can be calculated by
\begin{equation}
\int\frac{|\psi_{j}(\mathbf{r}')|^{2}}{1+|\mathbf{r}-\mathbf{r}'|^6} d\mathbf{r}'=\mathcal{F}^{-1}\left[\mathcal{F}\left[\frac{1}{1+\mathbf{r}^6}\right]\mathcal{F}\left[|\psi_{j}(\mathbf{r})|^{2}\right]\right],\tag{S4}
\end{equation}
where the Fourier transformation of the soft-core long-range potential is represented by the Meijer's $G$-function~\cite{Beals} as
\begin{equation}
\mathcal{F}\left[\frac{1}{1+\mathbf{r}^6}\right]=\frac{\pi}{3} G_{06}^{40}\left[\left(\frac{\mathbf{k}}{6}\right)^6\big|_{0,1/3,2/3,2/3,0,1/3}^{\text{ \ \ \ \ \ \ \ \ \ \ }-}\right].\tag{S5}
\end{equation}
In treating the time derivatives, we use the backward (forward) Euler scheme for the linear (nonlinear) terms, then the final discretization scheme is written as
\begin{subequations}
\begin{equation}
-\frac{\tilde{\psi}_{\uparrow}^{n+1}-\tilde{\psi}_{\uparrow}^{n}}{\tau}=\frac{\omega_{x}^2+\omega_{y}^2}{2}\tilde{\psi}_{\uparrow}^{n+1}+\kappa\left(
\omega_{x}-i\omega_{y}\right) \tilde{\psi}_{\downarrow}^{n+1}+\tilde{G}_{\uparrow}^{n},\tag{S6a}
\end{equation}
\begin{equation}
-\frac{\tilde{\psi}_{\downarrow}^{n+1}-\tilde{\psi}_{\downarrow}^{n}}{\tau}=\frac{\omega_{x}^2+\omega_{y}^2}{2}\psi_{\downarrow}^{n+1}+\kappa\left(
\omega_{x}+i\omega_{y}\right) \tilde{\psi}_{\uparrow}^{n+1}+\tilde{G}_{\downarrow}^{n}.\tag{S6b}
\end{equation}
\end{subequations}
A similar discretization scheme, named backward-forward Euler sine-pseudospectral discretization, has been proposed and demonstrated for Bose systems without SO coupling~\cite{Bao}.\newline

\textbf{Quantum depletion.} The validity of the Gross-Pitaevskii mean-field approximation can be checked by evaluating the quantum depletion caused by quantum fluctuations~\cite{Pethick}. According to the Bogoliubov theory, the fluctuation part $\delta\hat{\Psi}_{j}(\mathbf{r},t)$ with $(j=\uparrow, \downarrow)$ around the condensate can be subjected to a canonical transformation resulting in the expansion $\delta\hat{\Psi}_{j}(\mathbf{r},t)=\sum_{q}\big[u_{j q}(\mathbf{r}) e^{-i\omega_{q}t}\hat{\gamma}_{q}+v_{j q}^{\ast }(\mathbf{r}) e^{i\omega _{q}t} \hat{\gamma}_{q}^{\dag }\big]$, where $\hat{\gamma}_{q}$ and $\hat{\gamma}_{q}^{\dag}$ are the quasiparticle creation and annihilation operators associated with the $q$-th collective mode. The mode functions $u_{j q}(\mathbf{r})$, $v_{j q}(\mathbf{r})$ and the collective frequencies $\omega_{q}$ are determined by the Bogoliubov-de Gennes (BdG) equation
\begin{equation}
\begin{array}{l}
\left[
\begin{array}{cccc}
H_{s_{\uparrow }} & V_{\text{so}} &
0 & 0 \\
-V_{\text{so}}^{\ast } & H_{s_{\downarrow }} & 0 & 0 \\
0 & 0 &
H_{s_{\uparrow}} & V_{\text{so}}^{\ast } \\
0 & 0 &
-V_{\text{so}} & H_{s_{\downarrow }}%
\end{array}%
\right]\!\!\left[
\begin{array}{c}
u_{\uparrow q}\left( \mathbf{r}\right) \\
u_{\downarrow q}\left( \mathbf{r}\right) \\
v_{\uparrow q}\left( \mathbf{r}\right) \\
v_{\downarrow q}\left( \mathbf{r}\right)%
\end{array}%
\right]+\left[
\begin{array}{cccc}
g_{\uparrow \uparrow }\Psi _{\uparrow}\Psi _{\uparrow}^{\ast } & g_{\uparrow \downarrow }\Psi_{\uparrow }\Psi _{\downarrow }^{\ast } &
g_{\uparrow\uparrow }\Psi_{\uparrow }\Psi_{\uparrow } & g_{\uparrow \downarrow }\Psi _{\uparrow }\Psi _{\downarrow } \\
g_{\downarrow \uparrow }\Psi _{\downarrow}\Psi _{\uparrow }^{\ast } & g_{\downarrow\downarrow }\Psi _{\downarrow }\Psi _{\downarrow }^{\ast } & g_{\downarrow \uparrow }\Psi _{\downarrow }\Psi _{\uparrow } & g_{\downarrow \downarrow }\Psi_{\downarrow }\Psi_{\downarrow } \\
g_{\uparrow \uparrow }\Psi _{\uparrow }^{\ast}\Psi _{\uparrow }^{\ast} & g_{\uparrow\downarrow }\Psi _{\uparrow }^{\ast }\Psi _{\downarrow }^{\ast } &
g_{\uparrow \uparrow }\Psi _{\uparrow }^{\ast}\Psi _{\uparrow } & g_{\uparrow \downarrow }\Psi _{\uparrow }^{\ast}\Psi_{\downarrow } \\
g_{\downarrow \uparrow }\Psi _{\downarrow }^{\ast }\Psi _{\uparrow}^{\ast } & g_{\downarrow \downarrow }\Psi _{\downarrow }^{\ast}\Psi _{\downarrow }^{\ast} &
g_{\downarrow \uparrow }\Psi _{\downarrow }^{\ast }\Psi _{\uparrow} & g_{\downarrow \downarrow }\Psi_{\downarrow}^{\ast}\Psi_{\downarrow}%
\end{array}%
\right]\!\!\left[
\begin{array}{c}
u_{\uparrow q}\left( \mathbf{r}\right) \\
u_{\downarrow q}\left( \mathbf{r}\right) \\
v_{\uparrow q}\left( \mathbf{r}\right) \\
v_{\downarrow q}\left( \mathbf{r}\right)%
\end{array}%
\right] \\
+\int \!\left[
\begin{array}{cccc}
U_{\uparrow \uparrow }(\mathbf{r}\!-\!\mathbf{r}')\Psi _{\uparrow}\Psi _{\uparrow}'^{\ast } & U_{\uparrow \downarrow }(\mathbf{r}\!-\!\mathbf{r}')\Psi_{\uparrow }\Psi _{\downarrow }'^{\ast } &
U_{\uparrow\uparrow }(\mathbf{r}\!-\!\mathbf{r}')\Psi_{\uparrow }\Psi_{\uparrow }' & U_{\uparrow \downarrow }(\mathbf{r}\!-\!\mathbf{r}')\Psi _{\uparrow }\Psi _{\downarrow }' \\
U_{\downarrow \uparrow }(\mathbf{r}\!-\!\mathbf{r}')\Psi _{\downarrow}\Psi _{\uparrow }'^{\ast } & U_{\downarrow\downarrow }(\mathbf{r}\!-\!\mathbf{r}')\Psi _{\downarrow }\Psi _{\downarrow }'^{\ast } &
U_{\downarrow \uparrow }(\mathbf{r}\!-\!\mathbf{r}')\Psi _{\downarrow }\Psi _{\uparrow }' & U_{\downarrow \downarrow }(\mathbf{r}\!-\!\mathbf{r}')\Psi_{\downarrow }\Psi_{\downarrow }' \\
U_{\uparrow \uparrow }(\mathbf{r}\!-\!\mathbf{r}')\Psi _{\uparrow }^{\ast}\Psi _{\uparrow }'^{\ast} & U_{\uparrow\downarrow }(\mathbf{r}\!-\!\mathbf{r}')\Psi _{\uparrow }^{\ast }\Psi _{\downarrow }'^{\ast } &
U_{\uparrow \uparrow }(\mathbf{r}\!-\!\mathbf{r}')\Psi _{\uparrow }^{\ast}\Psi _{\uparrow }' & U_{\uparrow \downarrow }(\mathbf{r}\!-\!\mathbf{r}')\Psi _{\uparrow }^{\ast}\Psi_{\downarrow }' \\
U_{\downarrow \uparrow }(\mathbf{r}\!-\!\mathbf{r}')\Psi _{\downarrow }^{\ast }\Psi _{\uparrow}'^{\ast } & U_{\downarrow \downarrow }(\mathbf{r}\!-\!\mathbf{r}')\Psi _{\downarrow }^{\ast}\Psi _{\downarrow }'^{\ast} &
U_{\downarrow \uparrow }(\mathbf{r}\!-\!\mathbf{r}')\Psi _{\downarrow }^{\ast }\Psi _{\uparrow}' & U_{\downarrow \downarrow }(\mathbf{r}\!-\!\mathbf{r}')\Psi_{\downarrow}^{\ast}\Psi_{\downarrow}'%
\end{array}%
\right]\!\!\left[
\begin{array}{c}
u_{\uparrow q}\left( \mathbf{r}'\right) \\
u_{\downarrow q}\left( \mathbf{r}'\right) \\
v_{\uparrow q}\left( \mathbf{r}'\right) \\
v_{\downarrow q}\left( \mathbf{r}'\right)%
\end{array}%
\right] \!d\mathbf{r}'=\hbar\omega_{q}\!\left[
\begin{array}{c}
u_{\uparrow q}\left( \mathbf{r}\right) \\
u_{\downarrow q}\left( \mathbf{r}\right) \\
-v_{\uparrow q}\left( \mathbf{r}\right) \\
-v_{\downarrow q}\left( \mathbf{r}\right)%
\end{array}%
\right]
\end{array}\tag{S7}
\end{equation}
under the normalization $\int\big( \left\vert u_{\uparrow
q}\right\vert ^{2}+\left\vert u_{\downarrow q}\right\vert
^{2}-\left\vert v_{\uparrow q}\right\vert ^{2}-\left\vert
v_{\downarrow q}\right\vert ^{2}\big) d\mathbf{r}=1$. For simplicity, we have set $\Psi _{j }\equiv \Psi _{j }\left( \mathbf{r}\right)$, $\Psi _{j }'\equiv \Psi _{j }\left( \mathbf{r}'\right)$, $H_{s_{\uparrow }}=-\frac{\hbar ^{2}\mathbf{\nabla
}^{2}}{2M}+g_{\uparrow \uparrow
}\left\vert \Psi _{\uparrow }\right\vert ^{2}+g_{\uparrow \downarrow
}\left\vert \Psi _{\downarrow }\right\vert ^{2}+\int \big[U_{\uparrow\uparrow }(\mathbf{r}\!-\!\mathbf{r}')|\Psi_{\uparrow }'|^{2}+U_{\uparrow\downarrow }(\mathbf{r}\!-\!\mathbf{r}')|\Psi_{\downarrow }'|^{2}\big]d\mathbf{r}'-\mu_{\uparrow }$ and
$H_{s_{\downarrow }}=-\frac{\hbar ^{2}\mathbf{\nabla
}^{2}}{2M}+\left[g_{\downarrow
\downarrow }\left\vert \Psi _{\downarrow }\right\vert
^{2}+g_{\downarrow \uparrow }\left\vert \Psi _{\uparrow }\right\vert
^{2}\right]+\int d\mathbf{r}'\left[U_{\downarrow\downarrow }(\mathbf{r}\!-\!\mathbf{r}')|\Psi_{\downarrow }'|^{2}+U_{\downarrow\uparrow }(\mathbf{r}\!-\!\mathbf{r}')|\Psi_{\uparrow }'|^{2}\right]-\mu_{\downarrow}$ with $\mu_{\uparrow,\downarrow}$ the chemical potential, and $V_{\text{so}}=-\hbar(i \kappa_{x} \partial _{x}+\kappa_{y} \partial
_{y})$ in the case of Rashba SO coupling. At zero temperature, the number of the non-condensate
particles can be calculated by $\delta N=\int\sum_{q}\big(
\left\vert v_{\uparrow q}\right\vert ^{2}+\left\vert v_{\downarrow
q}\right\vert ^{2}\big) d\mathbf{r}$, where $q$ is restricted by the
nonnegative mode frequencies $\omega_{q}>0$.\newline
\begin{figure}[h!]
\centerline{\includegraphics[width=12cm,clip=]{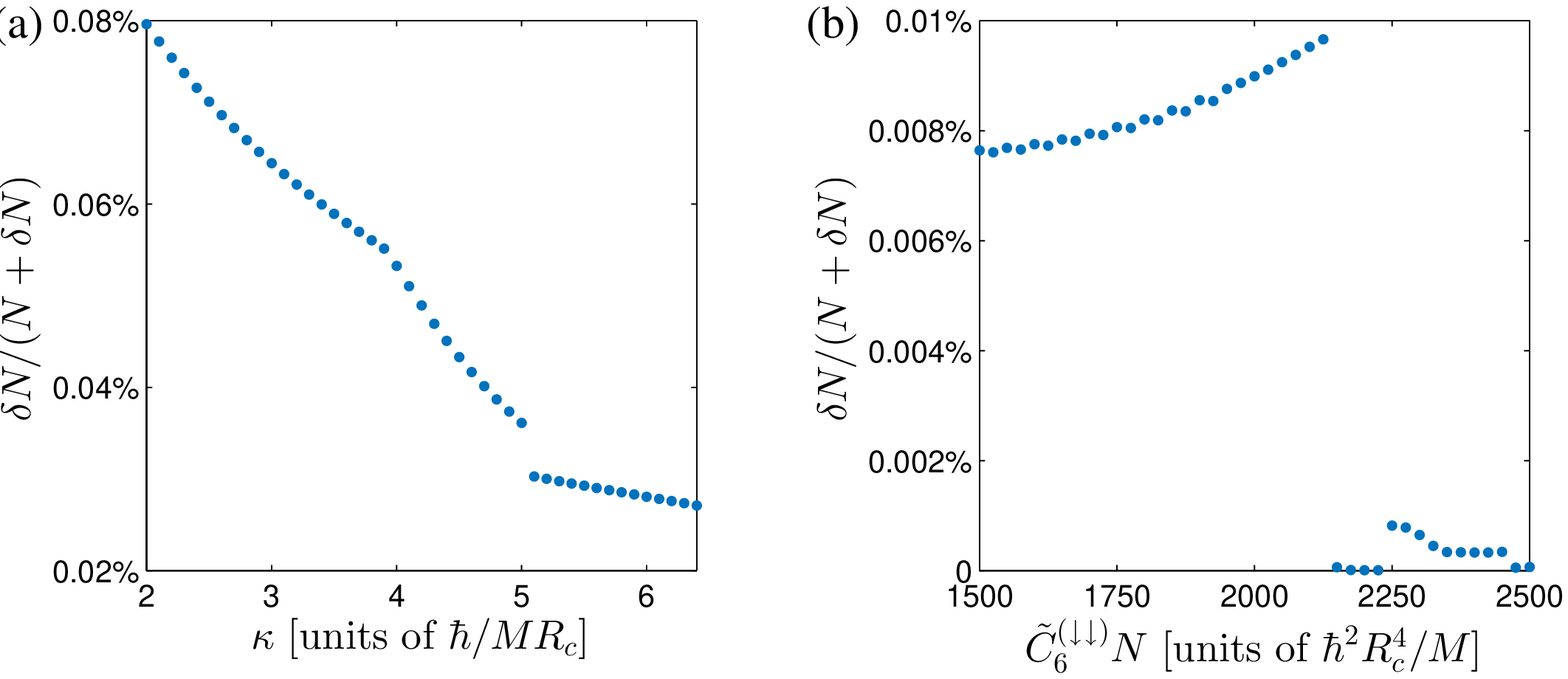}}
\setcounter{figure}{0}
\renewcommand{\thefigure}{S\arabic{figure}}
\caption{Quantum depletion as a function of (a) the
spin-orbit coupling strength $\kappa$ and (b) the soft-core long-range interaction strength $\tilde{C}_{6}^{(\downarrow\downarrow)}$. The soft-core long-range interaction strength is fixed at
$\tilde{C}_{6}^{(\downarrow\downarrow)}N=1300\ \hbar^2R^4_{c}/M$ in (a), and the spin-orbit coupling
strength is fixed at $\kappa=16\ \hbar/MR_{c}$ in (b). Other parameters are chosen as $\tilde{C}_{6}^{(\uparrow\uparrow)}N=2\tilde{C}_{6}^{(\uparrow\downarrow)}N=2500\ \hbar^2R^4_{c}/M$ and $gN=1000\ \hbar^2/M$. This set of parameters correspond to a system of about $N \sim 10^{5}$ particles with the $s$-wave scattering length $a_{ij}\approx 100 a_{B}$ with $a_{B}$ the Bohr radius.} \label{figS1}
\end{figure}
\begin{figure}[h!]
\centerline{\includegraphics[width=14cm,clip=]{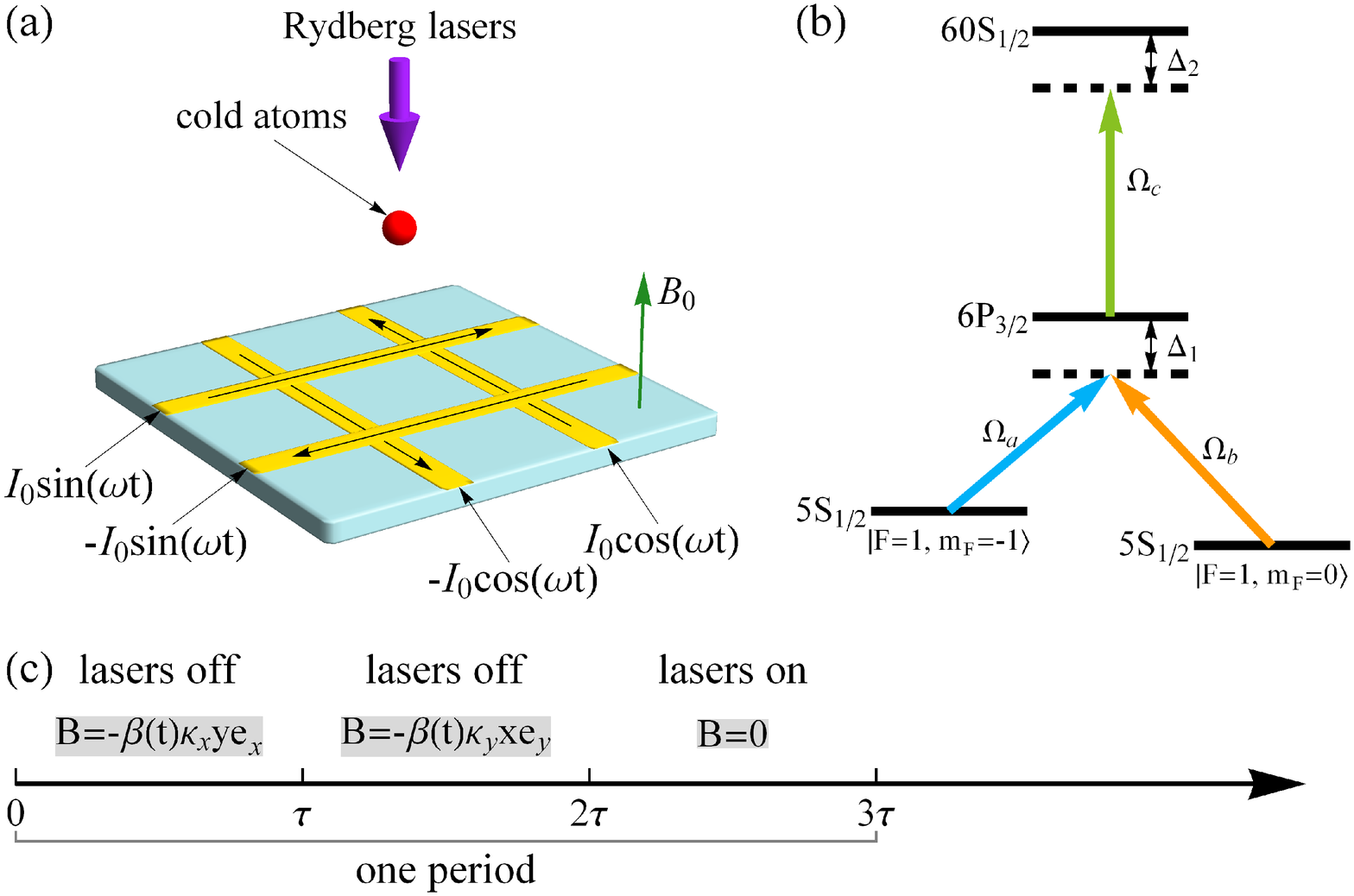}}
\setcounter{figure}{1}
\renewcommand{\thefigure}{S\arabic{figure}}
\caption{Scheme for creating Bose gases with spin-orbit coupling and soft-core long-range interactions. (a) Experimental setup. The
cloud of atoms is positioned tens of micrometers above the surface of
an atom chip. Two pairs of parallel microwires with amplitude
modulated radio-frequency current are embedded in the chip, and produce periodic
pulsed magnetic field gradients along perpendicular directions with the frequency equal to
the magnetic splitting induced by a strong bias field
$B_{0}\textbf{e}_{z}$ between the sublevels $m_{F}=-1$ and $m_{F}=0$. This induces an effective 2D spin-orbit coupling in the first-order approximation
to the pulse duration $\tau$. Three lasers are applied to coupled the two hyperfine ground states to a highly excited Rydberg state by a two-photon process, and create an effective soft-core long-rang potential in the condensate under the far-off-resonant condition. (b) Level diagram of the Rydberg dressing. (c) The pulse sequence used to implement spin-orbit coupling and soft-core long-range interaction. The parameters $\kappa_{x}$ and $\kappa_{y}$ characterize the strength of the magnetic field gradient, and
$\beta(t)$ defines the temporal shape of the magnetic fields. } \label{figS2}
\end{figure}

It is convenient to numerically solve the BdG equation~(S7) in the Fourier space. In Figs.~\ref{figS1}(a) and \ref{figS1}(b), we present the quantum depletion $\delta N/(N+\delta N)$ as a function of the SO coupling strength $\kappa$ and the soft-core interaction strength $\tilde{C}_{6}^{(\downarrow\downarrow)}$, respectively. One can see that, the quantum depletion is always less than $0.1\%$, thereby confirming the validity of the Gross-Pitaevskii mean-field approximation.\newline

\textbf{Experimental proposals.} The experimental realization of the supersolid phase shown in Fig. 1 of the main text may resort to the $^{\text{87}}$Rb platform, on which the Rydberg dressing technology~\cite{Pohl,WCWu,Heidemann} is used to creating the soft-core long-range interaction potential, and the Rashba SO coupling is created by modulating gradient magnetic field~\cite{ZFXu,BMAnderson,Ruquan-Wang}. The scheme is illustrated in Fig.~\ref{figS2}. Its feasibility relies on the ability to switch between magnetic pulses and laser pulses~[see Fig.~\ref{figS2}(c)].\newline

The first two stages of the scheme imitate the version of a recent proposal~\cite{BMAnderson}, where a strong time-independent bias magnetic field $\textbf{B}_{0}$ along the quantization axis $z$ and an infrared~(IR) magnetic field $\textbf{B}$ in the $x$-$y$ plane with a frequency $\omega$ in resonance with the splitting between the magnetic sublevels induced by the bias field~\cite{BMAnderson} are used to produce SO coupling between the two hyperfine states $\left\vert F=1,m_{F}=-1\right\rangle$ and $ \left\vert F=1,m_{F}=0\right\rangle$ of the $^{\text{87}}$Rb 5S$_{1/2}$ ground electronic manifold. The magnetic field with $\textbf{B}=-\beta(t)\kappa_{x}y\textbf{e}_{x}$ or $\textbf{B}=-\beta(t-\tau)\kappa_{y}x\textbf{e}_{y}$ is produced by a pair of wires along the $y$ or $x$ axis for the first ($0\leq t<\tau$) and the second ($\tau\leq t<2\tau$) stages, respectively~\cite{BMAnderson}, where $\kappa_{x,y}$ characterizes the strength of the magnetic field gradient, and $\beta(t)$ defines its temporal shape satisfying $\int_{0}^{\tau}\beta(t)dt=0$. This magnetic pulse imprints a spin-dependent phase gradient on the atoms, and produces a time-dependent effective Hamiltonian with the spin coupled to the atomic momentum~\cite{BMAnderson}. For a sufficiently short pulse period ($\tau\rightarrow0$), the time-averaged Hamiltonian for the first two stages takes an SO coupling term $\mathcal{V}_{\text{so}}=-i\hbar(\kappa_{x}\sigma_{x}\partial_{x}+\kappa_{y}\sigma_{y}\partial_{y})$ in the first-order approximation~\cite{BMAnderson,ZFXu}, where $\sigma_x$ and $\sigma_y$ are the quasi-spin operators for the selected pair of states. The SO-coupling parameters $\kappa_{x,y}$ depend on the strength of the magnetic pulses, and one arrives at the isotropic Rashba-type SO coupling for $\kappa_{x}=\kappa_{y}$.\newline

In the third stage, $2\tau\leq t<3\tau$, the the magnetic field is turned off and the Rydberg lasers are applied. Both the two hyperfine ground states $\left\vert F=1,m_{F}=-1\right\rangle$ and $ \left\vert F=1,m_{F}=0\right\rangle$ are optically coupled to the same excited $nS_{1/2}$ Rydberg state by a two-photon process, as shown in Fig.~S2(b). As a typical example, we choose $n=60$ and take $6P_{3/2}$ as the intermediate state. It is also supposed that the two hyperfine ground states experience different Rabi frequencies $\Omega_{a}$ and $\Omega_{b}$ of the first photon transition but the same detuning $\Delta_{1}$ and $\Delta_{2}$. In the region of a far-off-resonant coupling with $\Omega_{a,b}\ll \Delta_{1}$, the intermediate state $6P_{3/2}$ can be adiabatically eliminated, and the system reduces to an effective two-level atom with a two-photon Rabi frequency $\Omega_{\uparrow}=\frac{\Omega_{a}\Omega_{c}}{2\Delta_{1}}$ ($\Omega_{\downarrow}=\frac{\Omega_{b}\Omega_{c}}{2\Delta_{1}}$) and detuning $\Delta=\Delta_{1}+\Delta_{2}$ for the state $\left\vert F=1,m_{F}=-1\right\rangle$ ($ \left\vert F=1,m_{F}=0\right\rangle$)~\cite{Pohl}. In the case of weak coupling with $\Omega_{\uparrow,\downarrow}\ll \Delta$, the ground states are dressed by the effective soft-core long-range potential $U_{ij}\left(\mathbf{r}\right)=\tilde{C}_{6}^{(ij)}/\left(R_{c}^{6}+|\mathbf{r}|^6\right)$, where $\tilde{C}_{6}^{(ij)}=\frac{\Omega_{i}^2\Omega_{j}^2}{16\Delta^4}C_{6}$ ($i,j=\uparrow,\downarrow$) characterizes the interaction strength and $R_{c}=(C_{6}/2\hbar\Delta)^{1/6}$ represents the blockade radius with $C_{6}=9.7\times 10^{20}$ a.u. being the van der Waals coefficient of the $60S_{1/2}$ Rydberg state~\cite{Pohl,WCWu}. For a condensate with about $N=2\times 10^6$ atoms, the soft-core long-range interaction parameters used in Fig.~1 can be realized by adjusting the two-photon Rabi frequency $\Omega_{\uparrow}=420$ kHz, $\Omega_{\downarrow}=350$ kHz and detuning $\Delta=50$ MHz, which are experimentally accessible according the analysis of decoherence and loss mechanisms~\cite{Pohl}.

\textbf{Radius of the ring.} It has been known that the lattice constant of the supersolid induced by soft-core interaction $U_{ij}\left(\mathbf{r}\right)=\tilde{C}_{6}^{(ij)}/\left(R_{c}^{6}+|\mathbf{r}|^6\right)$ is mainly dominated by the Blockade radius $R_{c}$~\cite{Pohl}. As the Rydberg Blockade limits the size of the unit cells in the supersolid phase, the radius of the ring is less than $R_{c}$. In addition, it is also influenced by the interatomic interactions and the particle number distribution. While the radius increases with increasing hard-core interaction strength $g$, it decreases with increasing soft-core long-range interaction $\tilde{C}_{6}^{(ij)}$ [See Fig.~\ref{figS3}]. In the chiral supersolid (CSS) phase the particle number of the core component is large, and the ring's radius is relatively large. In contrast, in the anomalous chiral supersolid (ACSS) phase the particle number of the core component is small, and the the ring's radius is relatively small~[See Fig.~\ref{figS4}].\newline
\begin{figure*}[h!]
\centerline{\includegraphics[width=8.8cm,clip=]{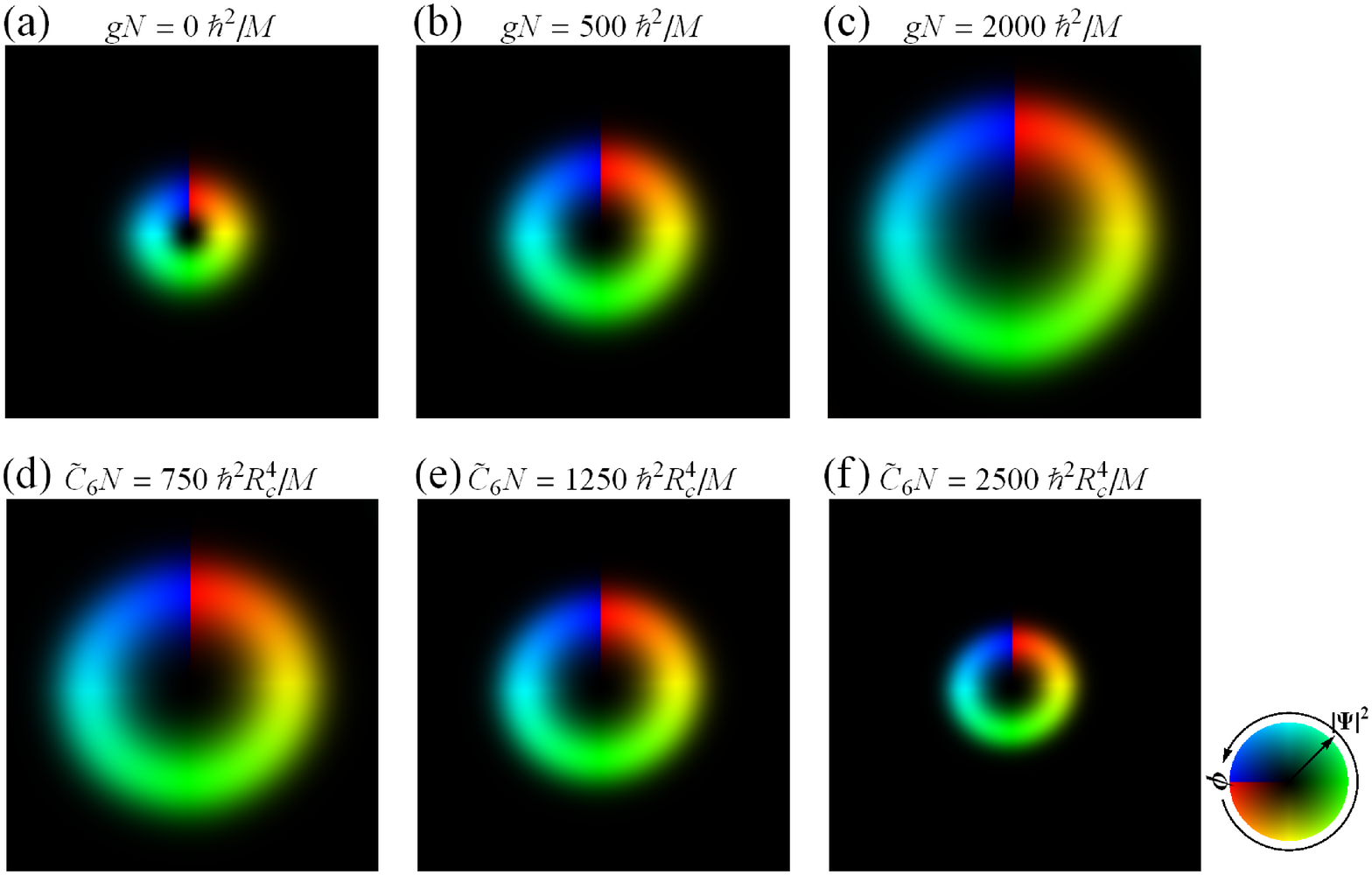}}
\setcounter{figure}{2}
\renewcommand{\thefigure}{S\arabic{figure}}
\caption{The toroidal component in a unit cell as a function of (a)-(c) the hard-core interaction strength $g$ and (d)-(f) the soft-core long-range interaction strength $\tilde{C}_{6}$. The density and phase distributions are represented by brightness and color, respectively. The soft-core long-range interaction strength in (a)-(c) is fixed at $\tilde{C}_{6}N=1250\ \hbar^2R^4_{c}/M$, and the hard-core interaction strength in (d)-(f) is fixed at $gN=1000\ \hbar^2/M$. The soft-core long-range interaction strengths satisfy $\tilde{C}_{6}=\tilde{C}_{6}^{(\downarrow\downarrow)}=\tilde{C}_{6}^{(\uparrow\downarrow)}=\tilde{C}_{6}^{(\uparrow\uparrow)}/2$. The spin-orbit coupling strength takes the value $\kappa=4\ \hbar/MR_{c}$. The size of each panel is $1.3 R_{c}$ along a side.}
\label{figS3}
\end{figure*}
\begin{figure*}[h!]
\centerline{\includegraphics[width=8.8cm,clip=]{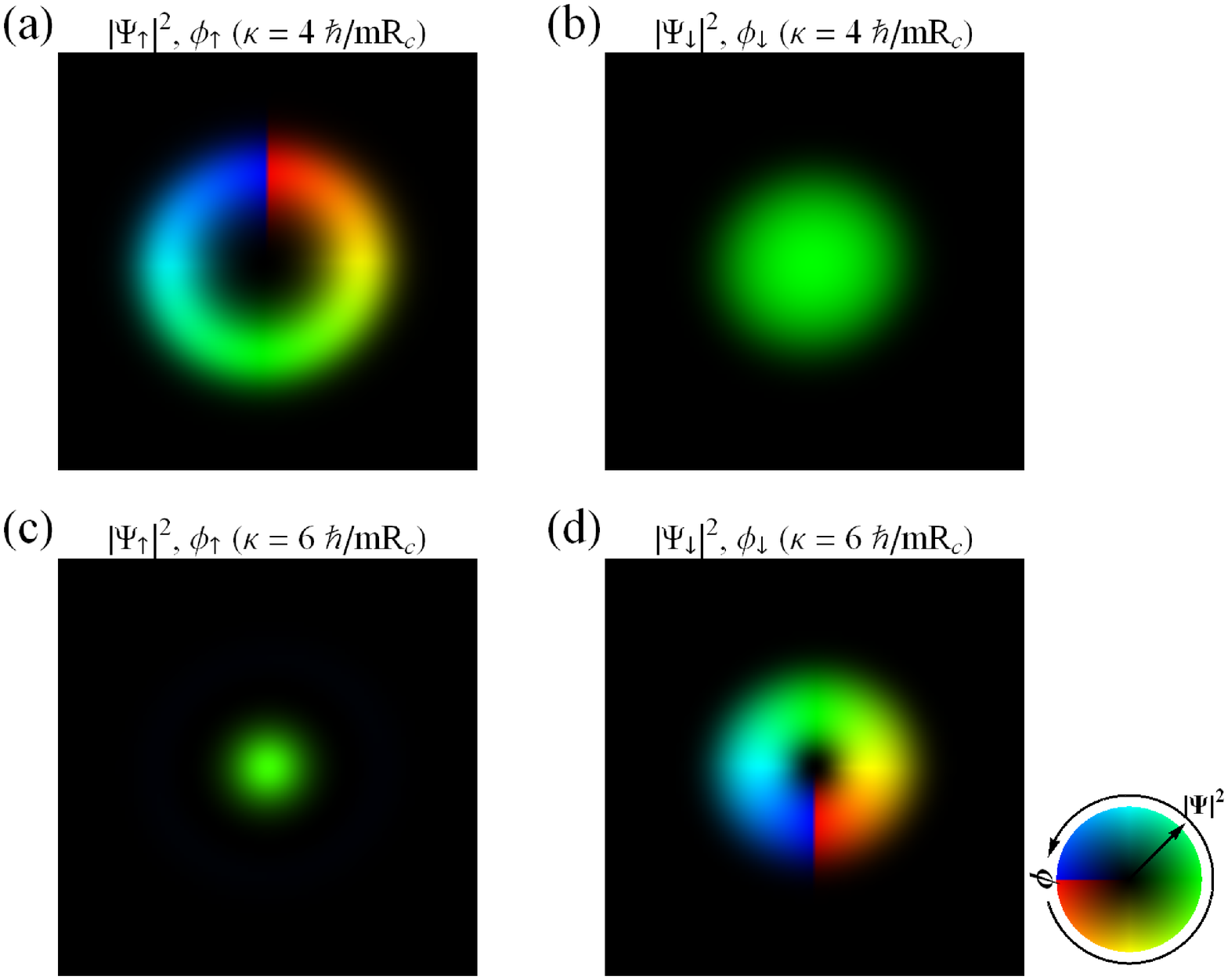}}
\setcounter{figure}{3}
\renewcommand{\thefigure}{S\arabic{figure}}
\caption{(a)-(b) One unit cell of the chiral supersolid (CSS) with weaker spin-orbit coupling $\kappa=4\ \hbar/MR_{c}$. (c)-(d) One unit cell of the anomalous chiral supersolid (ACSS) with stronger spin-orbit coupling $\kappa=6\ \hbar/MR_{c}$. The density and phase distributions are represented by brightness and color, respectively. The interatomic interactions are fixed at $\tilde{C}_{6}^{(\uparrow\uparrow)}N=2500\ \hbar^2R^4_{c}/M$, $\tilde{C}_{6}^{(\downarrow\downarrow)}N=\tilde{C}_{6}^{(\uparrow\downarrow)}N=1250\ \hbar^2R^4_{c}/M$ and $gN=1000\ \hbar^2/M$. The size of each panel is $1.3 R_{c}$ along a side.}
\label{figS4}
\end{figure*}

\textbf{Future prospects.} The SO-coupled Bose gas with soft-core long-range interactions investigated here is a multi-parameter complicated system. For simplicity, some of the parameters such as the spin-exchange interaction are ignored, and many other parameters including the hard-core interaction, the inter-component soft-core long-range interaction and the blockade radius are taken with fixed values. At the same time, the soft-core long-range interactions are supposed to be much stronger than the hard-core ones. A natural extension of our work is to investigate the ground-state phase diagram within the full parameter space, and discuss the influences of the not yet studied parameters. It should be predicted that the competition between the hard-core and soft-core long-range interactions, as well as the blockade radius, will lead to other exotic phenomena in the SO-coupled system.\newline

In addition to the Rashba and Dressehaus types, many new forms of SO coupling such as the NIST~\cite{Spielman}, Weyl~\cite{Spielman3}, Raman Lattice~\cite{Shuai-Chen} types as well as the angular SO coupling~\cite{Han-Pu,Chuanwei-Zhang} and SU(3) SO coupling~\cite{Galitski2} have been designed in hard-core systems, some of which has no analogue in condensed or any other forms of matters. The introduction of soft-core long-range interactions in these systems will open up a new avenue in cold atom physics, and make important breakthroughs in discovering novel states of matter and quantum phenomena.\newline\newline

\end{widetext}


\begin{thebibliography}{99}
\bibitem{McGuire} B. A. McGuire, P. B. Carroll, R. A. Loomis, I. A. Finneran, P. R. Jewell, A. J. Remijan, and G. A. Blake, Discovery of the interstellar chiral molecule propylene oxide (CH$_{3}$CHCH$_{2}$O), Science \textbf{352}, 1449 (2016).
\bibitem{Yoon} M. Yoon, R. Srirambalaji, and K. Kim, Homochiral metal-organic frameworks for asymmetric heterogeneous catalysis, Chem. Rev. \textbf{112}, 1196 (2012).
\bibitem{Kallin} C. Kallin and J. Berlinsky, Chiral superconductors, Rep. Prog. Phys. \textbf{79}, 054502 (2016).
\bibitem{Dai} H. Weng, C. Fang, Z. Fang, B. A. Bernevig, and X. Dai, Weyl Semimetal Phase in Noncentrosymmetric Transition-Metal Monophosphides, Phys. Rev. X \textbf{5}, 011029 (2015).
\bibitem{Parkin} K.-S. Ryu, L. Thomas, S.-H. Yang, and S. Parkin, Chiral spin torque at magnetic domain walls, Nat. Nanotechnol. \textbf{8}, 527 (2013).
\bibitem{Beach} S. Emori, U. Bauer, S.-M. Ahn, E. Martinez, and G. S. D. Beach, Current-driven dynamics of chiral ferromagnetic domain walls, Nat. Mater. \textbf{12}, 611 (2013).
\bibitem{Schmid} G. Chen, T. Ma, A. T. N'Diaye, H. Kwon, C. Won, Y. Wu, and A. K. Schmid, Tailoring the chirality of magnetic domain walls by interface engineering, Nat. Commun. \textbf{4}, 2671 (2013).
\bibitem{Tokura} K. Shibata, X. Z. Yu, T. Hara, D. Morikawa, N. Kanazawa, K. Kimoto, S. Ishiwata, Y. Matsui, and Y. Tokura, Towards control of the size and helicity of Skyrmions in helimagnetic alloys by spin-orbit coupling, Nat. Nanotechnol. \textbf{8}, 723 (2013).
\bibitem{Spielman} Y.-J. Lin, K. Jim\'{e}nez-Garc\'{\i}a, and I. B. Spielman, Spin-orbit-coupled Bose-Einstein condensates, Nature (London) \textbf{471}, 83 (2011).
\bibitem{Jing-Zhang} P. Wang, Z.-Q. Yu, Z. Fu, J. Miao, L. Huang, S. Chai, H. Zhai, and J. Zhang, Spin-Orbit Coupled Degenerate Fermi Gases, Phys. Rev. Lett. \textbf{109}, 095301 (2012).
\bibitem{Zwierlein} L. W. Cheuk, A. T. Sommer, Z. Hadzibabic, T. Yefsah, W. S. Bakr, and M. W. Zwierlein, Spin-Injection Spectroscopy of a Spin-Orbit Coupled Fermi Gas, Phys. Rev. Lett. \textbf{109}, 095302 (2012).
\bibitem{Jing-Zhang2} L. Huang, Z. Meng, P. Wang, P. Peng, S.-L. Zhang, L. Chen, D. Li, Q. Zhou, and J. Zhang, Experimental realization of two-dimensional synthetic spin-orbit coupling in ultracold Fermi gases, Nat. Phys. \textbf{12}, 540 (2016).
\bibitem{Shuai-Chen} Z. Wu, L. Zhang, W. Sun, X.-T. Xu, B.-Z. Wang, S.-C. Ji, Y. Deng, S. Chen, X.-J. Liu, and J.-W. Pan, Realization of two-dimensional spin-orbit coupling for Bose-Einstein condensates, Science \textbf{354}, 83 (2016).
\bibitem{JHHan} X.-Q. Xu and J. H. Han, Emergence of Chiral Magnetism in Spinor Bose-Einstein Condensates with Rashba Coupling, Phys. Rev. Lett. \textbf{108}, 185301 (2012).
\bibitem{SSZhang} S.-S. Zhang, W.-M. Liu, and H. Pu, Itinerant chiral ferromagnetism in a trapped Rashba spin-orbit-coupled Fermi gas, Phys. Rev. A \textbf{93}, 043602 (2016).
\bibitem{Galitski} V. Galitski and I. B. Spielman, Spin-orbit coupling in quantum gases, Nature (London) \textbf{494}, 49 (2013).
\bibitem{Dalibard} J. Dalibard, F. Gerbier, G. Juzeli\={u}nas, and P. \"{O}hberg, Colloquium: Artificial gauge potentials for neutral atoms, Rev. Mod. Phys. \textbf{83}, 1523 (2011).
\bibitem{Goldman} N. Goldman, G. Juzeli\={u}nas, P. \"{O}hberg, and I. B. Spielman, Light-induced gauge fields for ultracold atoms, Rep. Prog. Phys. \textbf{77}, 126401 (2014).
\bibitem{Hui-Zhai} H. Zhai, Spin-orbit coupled quantum gases, Int. J. Mod. Phys. B \textbf{26}, 1230001 (2012).
\bibitem{Hui-Zhai2} H. Zhai, Degenerate quantum gases with spin-orbit coupling: A review, Rep. Prog. Phys. \textbf{78}, 026001 (2015).
\bibitem{Yirev} W. Yi, W. Zhang, and X. L. Cui, Pairing superfluidity in spin-orbit coupled ultracold Fermi gases, Sci. China Phys. Mech. Astron. \textbf{58}, 014201 (2015).
\bibitem{Jing-Zhang3} J. Zhang, H. Hu, X. J. Liu, and H. Pu, ``Fermi gases with synthetic spin-orbit coupling" in Annual Review of Cold Atoms and Molecules Vol. 2, edited by K. Madison, K. Bongs, L. D. Carr, A. M. Rey, and H. Zhai (World Scientific, Singapore, 2014, p. 81.

     Annu. Rev. Cold At. Mol. \textbf{2}, 81 (2014).
\bibitem{Congjun-Wu} C. Wu, Unconventional Bose-Einstein condensations beyond the ``no-node" theorem, Mod. Phys. Lett. B \textbf{23}, 1 (2009).
\bibitem{Congjun-Wu2} C.-J. Wu, I. Mondragon-Shem, and X.-F. Zhou, Unconventional Bose-Einstein condensations from spin-orbit coupling, Chin. Phys. Lett. \textbf{28}, 097102 (2011).
\bibitem{Su-Yi} Y. Deng, J. Cheng, H. Jing, C.-P. Sun, and S. Yi, Spin-Orbit-Coupled Dipolar Bose-Einstein Condensates, Phys. Rev. Lett. \textbf{108}, 125301 (2012).
\bibitem{Clark} R. M. Wilson, B. M. Anderson, and C. W. Clark, Meron Ground State of Rashba Spin-Orbit-Coupled Dipolar Bosons, Phys. Rev. Lett. \textbf{111}, 185303 (2013).
\bibitem{Demler} S. Gopalakrishnan, I. Martin, and E. A. Demler, Quantum Quasicrystals of Spin-Orbit-Coupled Dipolar Bosons, Phys. Rev. Lett. \textbf{111}, 185304 (2013).
\bibitem{Pohl} N. Henkel, R. Nath, and T. Pohl, Three-Dimensional Roton Excitations and Supersolid Formation in Rydberg-Excited Bose-Einstein Condensates, Phys. Rev. Lett. \textbf{104}, 195302 (2010).
\bibitem{WCWu} C.-H. Hsueh, Y.-C. Tsai, K.-S. Wu, M.-S. Chang, and W. C. Wu, Pseudospin orders in the supersolid phases in binary Rydberg-dressed Bose-Einstein condensates, Phys. Rev. A \textbf{88}, 043646 (2013).
\bibitem{Heidemann} R. Heidemann, U. Raitzsch, V. Bendkowsky, B. Butscher, R. L\"{o}w, and T. Pfau, Rydberg Excitation of Bose-Einstein Condensates, Phys. Rev. Lett. \textbf{100}, 033601 (2008).
\bibitem{Boninsegni} M. Boninsegni and N. V. Prokof'ev, Colloquium: Supersolids: What and where are they? Rev. Mod. Phys. \textbf{84}, 759 (2012).
\bibitem{Boninsegni2} M. Boninsegni, Supersolid phases of cold atom assemblies, J. Low. Temp. Phys. \textbf{168}, 137 (2012).
\bibitem{Balibar} S. Balibar, The enigma of supersolidity, Nature \textbf{464}, 176 (2010).
\bibitem{Andreev} A. F. Andreev and I. M. Lifshitz, Quantum theory of defects in crystals, Zh. Eksp. Teor. Fiz. \textbf{56}, 2057 (1969) [Sov.-Phys. JETP \textbf{29}, 1107 (1969)].
\bibitem{Chester} G. V. Chester, Speculations on Bose-Einstein condensation and quantum crystals, Phys. Rev. A \textbf{2}, 256 (1970).
\bibitem{Leggett} A. J. Leggett, Can a Solid be ``Superfluid"?, Phys. Rev. Lett. \textbf{25}, 1543 (1970).
\bibitem{Kim} E. Kim and M. H. W. Chan, Probable observation of a supersolid helium phase, Nature (London) \textbf{427}, 225 (2004).
\bibitem{Bohm} D. Bohm, Note on a theorem of Bloch concerning possible causes of superconductivity, Phys. Rev. \textbf{75}, 502 (1949).
\bibitem{Momoi} Y. Ohashi and T. Momoi, On the Bloch theorem concerning spontaneous electric current, J. Phys. Soc. Jpn. \textbf{65}, 3254 (1996).
\bibitem{Dalibard2} K. W. Madison, F. Chevy, W. Wohlleben, and J. Dalibard, Vortex Formation in a Stirred Bose-Einstein Condensate, Phys. Rev. Lett. \textbf{84}, 806 (2000).
\bibitem{Ketterle} J. R. Abo-Shaeer, C. Raman, J. M. Vogels, and W. Ketterle, Observation of vortex lattices in Bose-Einstein condensates, Science \textbf{292}, 476 (2001).
\bibitem{Spielman2} Y.-J. Lin, R. L. Compton, K. Jim\'{e}nez-Garc\'{\i}a, J. V. Porto, and I. B. Spielman, Synthetic magnetic fields for ultracold neutral atoms, Nature (London) \textbf{462}, 628 (2009).
\bibitem{Supp} See Supplemental Material, which includes Refs.\cite{Spielman,Shuai-Chen,Pohl,WCWu,Heidemann,Dalfovo,Wei-Han,Beals,Bao,Pethick,ZFXu,BMAnderson,Ruquan-Wang,Spielman3,Han-Pu,Chuanwei-Zhang,Galitski2}, for more information about the numerical details, quantum depletion, experimental proposals, radius of the ring, and future prospects.
\bibitem{Dalfovo} F. Dalfovo and S. Stringari, Bosons in anisotropic traps: Ground state and vortices, Phys. Rev. A \textbf{53}, 2477 (1996); M. L. Chiofalo, S. Succi, and  M. P. Tosi, Ground state of trapped interacting Bose-Einstein condensates by an explicit imaginary-time algorithm, Phys. Rev. E \textbf{62}, 7438 (2000).
\bibitem{Wei-Han} W. Han, G. Juzeli\={u}nas, W. Zhang, and W.-M. Liu, Supersolid with nontrivial topological spin textures in spin-orbit-coupled Bose gases, Phys. Rev. A \textbf{91}, 013607 (2015).
\bibitem{Beals} R. Beals and J. Szmigielski, Meijer G-functions: A gentle introduction, Not. Am. Math. Soc. \textbf{60}, 866 (2013).
\bibitem{Bao} W. Bao, I.-L. Chern, and F. Y. Lim, Efficient and spectrally accurate numerical methods for computing ground and first excited states in Bose-Einstein condensates, J. Comput. Phys. \textbf{219}, 836 (2006).
\bibitem{Pethick} C. J. Pethick and H. Smith, {\it Bose-Einstein Condensation in Dilute Gases} (Cambridge University Press, Cambridge, England, 2002).
\bibitem{ZFXu} Z.-F. Xu, L. You, and M. Ueda, Atomic spin-orbit coupling synthesized with magnetic-field-gradient pulses, Phys. Rev. A \textbf{87}, 063634 (2013).
\bibitem{BMAnderson} B. M. Anderson, I. B. Spielman, and G. Juzeli\={u}nas, Magnetically Generated Spin-Orbit Coupling for Ultracold Atoms, Phys. Rev. Lett. \textbf{111}, 125301 (2013).
\bibitem{Ruquan-Wang} X. Luo, L. Wu, J. Chen, Q. Guan, K. Gao, Z.-F. Xu, L. You, and R. Wang, Tunable atomic spin-orbit coupling synthesized with a modulating gradient magnetic field, Sci. Rep. \textbf{6}, 18983 (2016).
\bibitem{Spielman3} B. M. Anderson, G. Juzeli\={u}nas, V. M. Galitski, and I. B. Spielman, Synthetic 3D Spin-Orbit Coupling, Phys. Rev. Lett. \textbf{108}, 235301 (2012).
\bibitem{Han-Pu} M. DeMarco and H. Pu, Angular spin-orbit coupling in cold atoms, Phys. Rev. A \textbf{91}, 033630 (2015).
\bibitem{Chuanwei-Zhang} K. Sun, C. Qu, and C. Zhang, Spin-orbital-angular-momentum coupling in Bose-Einstein condensates, Phys. Rev. A \textbf{91}, 063627 (2015).
\bibitem{Galitski2} R. Barnett, G. R. Boyd, and V. Galitski, SU(3) Spin-Orbit Coupling in Systems of Ultracold Atoms, Phys. Rev. Lett. \textbf{109}, 235308 (2012).
\bibitem{Campbell} D. L. Campbell, G. Juzeli\={u}nas, and I. B. Spielman, Realistic Rashba and Dresselhaus spin-orbit coupling for neutral atoms. Phys. Rev. A \textbf{84}, 025602 (2011).
\bibitem{Yefsah} T. Yefsah, R. Desbuquois, L. Chomaz, K. J. G\"{u}nter, and J. Dalibard, Exploring the Thermodynamics of a Two-Dimensional Bose Gas, Phys. Rev. Lett. \textbf{107}, 130401 (2011).
\bibitem{Santos} S. Sinha, R. Nath, and L. Santos, Trapped Two-Dimensional Condensates with Synthetic Spin-Orbit Coupling, Phys. Rev. Lett. \textbf{107}, 270401 (2011).
\bibitem{Hu} H. Hu, B. Ramachandhran, H. Pu, and X.-J. Liu, Spin-Orbit Coupled Weakly Interacting Bose-Einstein Condensates in Harmonic Traps, Phys. Rev. Lett. \textbf{108}, 010402 (2012).
\bibitem{ZFXu2} Z. F. Xu, R. L\"{u}, and L. You, Emergent patterns in a spin-orbit-coupled spin-2 Bose-Einstein condensate, Phys. Rev. A \textbf{83}, 053602 (2011).
\bibitem{SCGou} S.-W. Su, I.-K. Liu, Y.-C. Tsai, W. M. Liu, and S.-C. Gou, Crystallized half-Skyrmions and inverted half-Skyrmions in the condensation of spin-1 Bose gases with spin-orbit coupling, Phys. Rev. A \textbf{86}, 023601 (2012).
\bibitem{Ueda} Z. F. Xu, Y. Kawaguchi, L. You, and M. Ueda, Symmetry classification of spin-orbit-coupled spinor Bose-Einstein condensates, Phys. Rev. A \textbf{86}, 033628 (2012).
\bibitem{Mottonen} E. Ruokokoski, J. A. M. Huhtam\"{a}ki, and M. M\"{o}tt\"{o}nen, Stationary states of trapped spin-orbit-coupled Bose-Einstein condensates, Phys. Rev. A \textbf{86}, 051607(R) (2012).
\bibitem{Ueda2} Z.-F. Xu, S. Kobayashi, and M. Ueda, Gauge-spin-space rotation-invariant vortices in spin-orbit-coupled Bose-Einstein condensates, Phys. Rev. A \textbf{88}, 013621 (2013).
\bibitem{SCGou2} S.-W. Su, S.-C. Gou, Q. Sun, L. Wen, W.-M. Liu, A.-C. Ji, J. Ruseckas, and G. Juzeli\={u}nas, Rashba-type spin-orbit coupling in bilayer Bose-Einstein condensates, Phys. Rev. A \textbf{93}, 053630 (2016).
\bibitem{YTokura} N. Nagaosa and Y. Tokura, Topological properties and dynamics of magnetic Skyrmions, Nat. Nanotechnol. \textbf{8}, 899 (2013).
\bibitem{CPfleiderer} U. K. R\"{o}{\ss}ler, A. N. Bogdanov, and C. Pfleiderer, Spontaneous Skyrmion ground states in magnetic metals, Nature (London) \textbf{442}, 797 (2006).
\bibitem{CPfleiderer2} S. M\"{u}hlbauer, B. Binz, F. Jonietz, C. Pfleiderer, A. Rosch, A. Neubauer, R. Georgii, and P. B\"{o}ni, Skyrmion lattice in a chiral magnet, Science \textbf{323}, 915 (2009).
\bibitem{YTokura2} X. Z. Yu, Y. Onose, N. Kanazawa, J. H. Park, J. H. Han, Y. Matsui, N. Nagaosa, and Y. Tokura, Real-space observation of a two-dimensional Skyrmion crystal, Nature (London) \textbf{465}, 901 (2010).
\bibitem{YTokura3} S. Seki, X. Z. Yu, S. Ishiwata, and Y. Tokura, Observation of Skyrmions in a multiferroic material, Science \textbf{336}, 198 (2012).
\bibitem{Baym} T. Ozawa and G. Baym, Striped states in weakly trapped ultracold Bose gases with Rashba spin-orbit coupling, Phys. Rev. A \textbf{85}, 063623 (2012).
\bibitem{Fetter} A. L. Fetter, Vortex dynamics in spin-orbit-coupled Bose-Einstein condensates, Phys. Rev. A \textbf{89}, 023629 (2014).
\bibitem{Note01} The left-hand (right-hand) rule indicates that the left (right) fingers are curled in the direction of circulation and the left (right) thumb points in the spin orientation in the vortex core.
\bibitem{Note02} Even in the case of $\tilde{C}_{6}^{(\uparrow\uparrow)}=\tilde{C}_{6}^{(\downarrow\downarrow)}$, the absolute majority of particles still populate the core component, leading to a spontaneously emerged population imbalance.
\bibitem{Hui-Zhai3} C. Wang, C. Gao, C.-M. Jian, and H. Zhai, Spin-Orbit Coupled Spinor Bose-Einstein Condensates, Phys. Rev. Lett. \textbf{105}, 160403 (2010).
\bibitem{Ketterle2} J.-R. Li, J. Lee, W. Huang, S. Burchesky, B. Shteynas, F. \c{C}. Top, A. O. Jamison, and W. Ketterle, A stripe phase with supersolid properties in spin-orbit-coupled Bose-Einstein condensates, Nature (London) \textbf{543}, 91 (2017).
\bibitem{Ho} T.-L. Ho and S. Zhang, Bose-Einstein Condensates with Spin-Orbit Interaction, Phys. Rev. Lett. \textbf{107}, 150403 (2011).
\bibitem{Shuai-Chen2} S.-C. Ji, J.-Y. Zhang, L. Zhang, Z.-D. Du, W. Zheng, Y.-J. Deng, H. Zhai, S. Chen, and J.-W. Pan, Experimental determination of the finite-temperature phase diagram of a spin-orbit coupled Bose gas, Nat. Phys. \textbf{10}, 314 (2014).
\bibitem{Stringari} Y. Li, L. P. Pitaevskii, and S. Stringari, Quantum Tricriticality and Phase Transitions in Spin-Orbit Coupled Bose-Einstein Condensates, Phys. Rev. Lett. \textbf{108}, 225301 (2012).
\bibitem{Cornell} D. S. Hall, M. R. Matthews, J. R. Ensher, C. E. Wieman, and E. A. Cornell, Dynamics of Component Separation in a Binary Mixture of Bose-Einstein Condensates, Phys. Rev. Lett. \textbf{81}, 1539 (1998).
\end{thebibliography}
\end{document}